\documentclass[aps, draft, floats,showpacs,nofootinbib]{revtex4}
\usepackage{epsf}
\usepackage{graphicx}
\usepackage{color}
\definecolor{orange}{rgb}{1,.7,0}
\def\chg#1{#1}
\def\ical{{\cal I}}
\def\lcal{{\cal L}}
\def\ocal{{\cal O}}
\def\fcal{{\cal F}}
\def\inv#1{{1\over #1}}
\def\half{\inv2}
\def\tr{{\hbox{Tr}}}
\def\rhs{right hand side\ }
\def\taubf{\mbox{\boldmath$\tau$}}
\def\pibf{\mbox{\boldmath$\pi$}}
\def\gesim{\,{\raise-3pt\hbox{$\sim$}}\!\!\!\!\!{\raise2pt\hbox{$>$}}\,}
\def\lesim{\,{\raise-3pt\hbox{$\sim$}}\!\!\!\!\!{\raise2pt\hbox{$<$}}\,}
\def\sqr#1#2{{\vcenter{\hrule height.#2pt
      \hbox{\vrule width.#2pt height#1pt \kern#1pt
         \vrule width.#2pt}
      \hrule height.#2pt}}}
\def\square{\mathchoice\sqr56\sqr56\sqr{2.1}3\sqr{1.5}3}
\def\up#1{^{\left( #1 \right) }}
\def\Xb{{\overline{X}}}
\newcommand{\nc}{\newcommand}
\nc{\beq}{\begin{equation}}  \nc{\eeq}{\end{equation}}
\nc{\bea}{\begin{eqnarray}}  \nc{\eea}{\end{eqnarray}}

\begin{document}

\title{ On the Vafa-Witten Theorem on Spontaneous Breaking of Parity}

\author{Martin B. Einhorn}
\affiliation{Michigan Center for Theoretical Physics,\\ 
	     Randall Laboratory, The University of Michigan, Ann Arbor, MI 48109\\
	     E-mail: {\tt meinhorn@umich.ed}}
\author{Jos\'e Wudka}
\affiliation{Department of Physics\\
	     University of California, Riverside CA 92521-0413, USA\\
	     E-mail: {\tt Jose.Wudka@cur.ed}}

\date{\today}

\begin{abstract}
We revisit the proof provided by Vafa and Witten that there is no
spontaneous parity breaking in theories with vector-like fermions.
We argue that the various criticisms that have been leveled at the
original proof do not invalidate it
\end{abstract}


\preprint{MCTP-02-23}
\preprint{UCRHEP-T337}

\maketitle

\section{Introduction}
\label{sect:introduciton}

Several years ago, Vafa and Witten\cite{vw} (hereafter
referred to as VW) provided a simple, elegant 
argument for the absence of spontaneous parity violation
in theories containing only vector-like fermions; this result,
however, has been criticized on various grounds
dealing with mathematical niceties underlying the proof.
It was argued in~\cite{a.g} and later in~\cite{j}
that the expression for the free 
energy on which the VW argument is based has an ill-defined 
infinite-volume limit and is therefore insufficient to determine
the absence of spontaneous parity violation.
Previously, Aoki\cite{aoki} 
had noted the possibility that lattice QCD
has a parity-violating phase which disappears in the continuum limit;
this apparently presents a problem for the VW result since the original proof 
can be applied to the lattice-regulated theory as well as to the 
continuum. Hence the Aoki phase, even if only present at non-zero lattice
spacing, would directly contradict the results of~\cite{vw}
{\em provided} the assumptions made by VW are satisfied by the
lattice regulated theory at finite spacing. This model
was studied carefully by Sharpe and Singleton\cite{ss} who
argued that the Aoki phase is not ruled out by the results of VW
since their arguments appear not to be applicable for
the case of fermionic order parameters. 
Finally, reference~\cite{cohen} raised the
possibility of parity-violating operators that evade the proof 
of VW for systems in a thermal bath.  In this paper, we will 
re-examine these complaints and find that, 
for physically relevant systems, they do not invalidate the results of 
VW.\cite{vw}In a related paper,\cite{ew} we argue that the proof, 
 even under the assumptions stated, does not however exclude 
spontaneous parity violation under certain circumstances.

\medskip

We organize this paper as follows. In section
\ref{sect:proof} we review the original VW argument. 
In section \ref{sect:large.v} we examine
the large volume limit of the theory and show that, as claimed
in~\cite{a.g}, it is well defined only for the case where
spontaneous symmetry breaking does not occur; yet we will
argue that this deficiency does 
not invalidate the results of VW. We revisit the
lattice arguments in section \ref{sect:lattice} where we argue
that the Aoki phases (at least for a large number of fermions)
do not represent absolute minima of the
effective potential but only metastable states~\footnote{This
should not be used to completely dismiss these states as irrelevant:
recent experiments in heavy-ion collisions create finite-volume
systems that may relax temporarily into parity-violating 
configurations\cite{br}}. In section \ref{sect:fermion}
we consider the use of fermionic operators as order parameters for a
possible parity-violating phase transition; in contrast to
the results of~\cite{ss} we find that this type of order
parameter exhibits the same behavior claimed by VW.
We then argue that there is a simple and useful fermionic
operator that can be better used as an order parameter
since it preserves chirality (the operators used in~\cite{aoki,ss}
violate both parity and chiral symmetry). Parity violation at
finite temperature is briefly revisited in Sect.~\ref{sect:temp}, and
parting comments are presented in section \ref{sect:conclusions}. A
mathematical result is relegated to the Appendix.

\section{The Vafa-Witten argument}
\label{sect:proof}
In this section we summarize the original VW argument. The
 idea is to consider a theory with action $ S $
that is even under parity transformations and contains only
vector-like fermions (so that the corresponding fermionic
determinant is real), and to add a parity-violating term
$X$ so that the full Minkowski-space action becomes
\beq
S_{\rm tot} = S + \lambda X,
\label{full.s}
\eeq
where $\lambda $ is a small real parameter and $X$ is
Hermitian. We will assume (without loss of generality) that $X$
is parity odd; all gauge-fixing terms are assumed to
be contained in $ S $. The bosonic terms in $S$ are
assumed to be real (this is always possible for
all linear gauges)~\footnote{For this
it is convenient {\em not} to introduce ghosts and deal with the
gauge-fixing determinant directly.}, 
(see for example~\cite{iz}). We will concentrate
on theories that do not contain scalar fields.

We first consider the case where 
$X$ is a purely bosonic operator. Some comments concerning scalar theories
are provided in section \ref{sect:conclusions}. We first obtain the
free energy by Wick-rotating $S$  in (\ref{full.s}). 
While not without ambiguities, to be precise, we will follow 
the procedure of~\cite{v.n}. Any parity-violating operator
contains an $\epsilon$ tensor and this implies that the corresponding
term acquires an additional factor of $i$ upon Wick 
rotation, thus we find
\beq
i S_{\rm tot} \to - S_E + i \lambda X_E,
\eeq
with $S_E$ and $X_E$ real (the subscript $E$ will be used
to denote the quantities in Euclidean space). The ``free energy" $W$ is then
\beq
e^{-W(\lambda) } = \int [dA]
e^{-S_E^{\rm boson} + i \lambda X_E} \;
{\rm Det}(\not\!\!D_E + M ),
\label{def.of.W}
\eeq
where $A$ denotes the gauge fields, and $S_E^{\rm boson} $ the purely 
bosonic contributions to $S_E$.\footnote{For constant $\lambda$, 
$W(\lambda)$ will be proportional to the space-time volume.} We take the mass
matrix $M$ diagonal with non-negative eigenvalues
(as we are not interested in explicit P violation); the 
Euclidean-space fermionic covariant derivative
is denoted by $D_E$ which is anti-Hermitian in our 
conventions~\cite{v.n} (the gamma  matrices are Hermitian). The 
fermionic determinant is then a real functional of the gauge fields
for vector-like couplings\cite{real.det}.

Consider now the case where $X$ is a fermionic bilinear~\footnote{When
$X$ contains products of bilinears these can be turned into
a sum of bilinears through the introduction of appropriate auxiliary
fields by using the Hubbard-Stratonovich trick~\cite{h.s}; the arguments
below carry through provided these auxiliary fields are vectors;
this will be investigated in detail in~\cite{ew}.},
\beq
X= \overline{ \psi} \ocal \psi.  
\eeq
After a Wick rotation and
to first order in $ \lambda $ the fermionic integration yields
\beq
e^{-W(\lambda) } = \int [dA] e^{-S_E^{\rm boson} + 
i \lambda X^{(\rm{eff})}_E }{\rm Det}(\not\!\!D_E + M ); \quad
X^{(\rm{eff})}_E = -i \hbox{Tr} \left[{1\over \not\!\! D_E + M} \ocal_E \right],
\eeq
where $ \ocal_E = \gamma^0 \gamma^5 \ocal \gamma^5 \gamma^0 $~\footnote{It
is understood that $ \ocal_E $ is obtained by replacing $D_0 \to
- i D_0^E,~\gamma^0 \to \gamma_0^E,~\gamma^k \to - i \gamma_k^E $ in $ \ocal $.
In our conventions $ \gamma_\mu^E{}^\dagger = + \gamma_\mu^E $ and
$ \{ \gamma_\mu^E,  \gamma_\nu^E \} = \delta_{\mu\nu} $}. 
A naive calculation shows that $ X^{(\rm{eff})}_E $ is real if
$ \ocal_E^\dagger = - \gamma^5 \ocal_E \gamma^5 $.
In particular note that this holds for the simple
cases $ \ocal_E = i \gamma_5 $ and
$ \ocal_E = {\buildrel \leftrightarrow \over {\not\!\!D}}_E\gamma^5 $. 
There is, of course, a possibility that 
these rather formal results are flawed by the mathematical subtleties
involved in properly defining $ X^{(\rm{eff})}_E $\cite{ss}; 
we discuss this possibility it in section
\ref{sect:fermion}.

With these preliminaries the proof is based on the claim~\cite{vw} 
that all parity-violating operators correspond to real
$ X_E$ or $ X_E^{(\rm eff)} $, in which case
\beq
W(\lambda) \ge W(0),
\label{bound}
\eeq
as a result of Schwartz's inequality. The validity of this result
then rests on the requirement that the measure is positive definite,
that is, ${\rm Det}(\not\!\!D_E + M ) > 0 $. This is true for the
case of vector-like fermions~\cite{rfd}, but can be extended to
some more general types of fermionic Lagrangians; we return to this
point in section \ref{sect:lattice}. Note that in proving (\ref{bound})
we used the fact that $W$ is real since $X$ is odd under
parity and the integration involves both a configuration and its parity conjugate. 

Since any explicit violation of the symmetry (in this case parity) 
results in an increased  free energy, the lowest energy ground state 
will be parity symmetric.  To see that eq.~(\ref{bound}) implies vanishing 
of the expectation value of $X$, $\Xb$, it is useful 
to pass to the effective potential $V(\Xb)$
\beq
V(\Xb)\equiv W(\lambda)- \lambda \Xb,\ \ {\rm with}\ 
\ \Xb \equiv \frac{\partial W}{\partial\lambda}.
\label{veff}
\eeq
Inversely, using $\lambda = -V'(\Xb)$, we can write this as 
\beq
W(\lambda)=V(\Xb) + \lambda \Xb = V(\Xb) - \Xb V'(\Xb). 
\label{veff2}
\eeq
We want to show that, in the limit $\lambda\!\to\!0^\pm,$ 
$\Xb \!\to\!0.$  Suppose, on the contrary, 
that~\footnote{The case $ \lambda \to 0^-$ is similar.}
as $ \lambda \to 0^+$, $\Xb = v \ne 0.$  
This is necessarily at a minimum 
of $V(\Xb)$.\footnote{Since the action $S_E$ is parity symmetric, 
and $\Xb$ is parity odd, $W(\lambda)=-W(-\lambda),$ and $V(\Xb)=V(-\Xb).$  
So, $-v$ will also be a minimum, but we simply pick one of them.}  
If we adopt the usual convention that $W(0)=0$, then $V(v)=0$ as well.  
The inequality eq.~(\ref{bound}) therefore implies that 
\beq
V(\Xb) - \Xb V'(\Xb)=  \left\{ V(\Xb)-(\Xb-v)V'(\Xb) \right\}-v V'(\Xb) \ge 0.
\label{bound2}
\eeq
Since $v$ is a minimum, $V'(v)=0$.  If we expand in a neighborhood of 
$\Xb=v$, the term in the curly brackets contains at least one power of 
$(\Xb-v)$ greater than the last term and, moreover, since $v$ is a minimum, 
the leading term in the expansion of $V'(\Xb)$ will be an odd power of 
$(\Xb-v)$.  (For $V''(v)>0,$ it will be linear in $(\Xb-v).$)  But then 
the inequality will be violated in a neighborhood of $\Xb=v.$  Therefore, 
the only possibility is that $v=0.$~\footnote{The same result
is obtained when one defines the effective potential in the region 
$ |\Xb|< |v| $ using the Maxwell construction~\cite{stan}.}.

The same conclusion can be obtained starting form the
Hamiltonian derived form (\ref{full.s})~\footnote{The following is
a clarification of an argument presented in \cite{vw}}. Indeed, 
if the ground state is not parity symmetric, then, when $ \lambda =0 $, 
there will be two degenerate lowest-energy states $ | \pm \rangle $ 
that are parity conjugate of each other. Now, the Hamiltonian 
corresponding to (\ref{full.s}) is of the form $ H + \lambda \Xi $ 
where $H$ is parity even and $ \Xi $ parity odd; a straightforward
application of degenerate perturbation theory shows that the term 
$ \lambda \Xi $ lifts the degeneracy, so that the lowest energy 
state for the system equals 
\beq
E_0 - \left| \lambda \epsilon \right| + O(\lambda^2 ) ;
\qquad E_0 =\langle \pm | H   | \pm \rangle ;~
  \epsilon = \langle +  | \Xi |  -   \rangle 
\eeq
that {\em decreases} as a function of $ \lambda $.  This contradicts 
(\ref{bound}) and implies that there are no parity-degenerate vacua: 
the symmetry is not broken.

\section{Large volume effects}
\label{sect:large.v}

The results summarized in the previous section were criticized
in~\cite{a.g} based on the following observation. Suppose we
construct a Landau-Ginzburg functional of the order parameter defined
as $ X_E/V $, where $V$ denotes the spatial volume of the system.
Explicitly,
\bea
e^{-W } &=&  \int_{-\infty}^\infty d\rho 
\int [dA] e^{-S_E^{\rm boson} + i \lambda X_E} \;
{\rm Det}(\not\!\!D_E + M ) \delta( \rho - X_E/V), \cr
&=&  \int_{-\infty}^\infty d\rho e^{-\fcal(\rho) + i V \lambda \rho},
\label{large.V.W}
\eea
where $ \fcal $ is presumably proportional to $V$ and
we assume it can be expanded in a power series in $ \rho $.
Since the original action is parity symmetric $ \fcal $ will
be even in $\rho $. For definiteness we will consider the
case
\beq
\fcal(\rho) = V \left[ \inv4 b \rho^4 + \inv2 a \rho^2 \right],
\eeq
hence
\bea
e^{-W} &=& (V b)^{-1/4} \; \ical(q) \cr
\ical(q) &=& \int_{-\infty}^\infty dx\;
\exp\left[ - \inv4 x^4 - \alpha q^{2/3} x^2 + i q x \right]
\eea
where
\beq
q = \lambda \left( V^3/b \right)^{1/4}; \qquad 
\alpha = \inv2 a \left( \lambda^2 b \right)^{-1/3}.
\eeq
The integral $ \ical $ is evaluated in the Appendix, 
we find, for small $|\lambda|$,
\bea
\ical
&\simeq& z e^{- V [a^2/(4b) + \lambda^2/(4a) + \cdots]}
 \cos \left(V |\lambda| \sqrt{|a|/b}+ \phi \right) ; 
\quad |\lambda| \ll 1,~a<0\cr
&\simeq& z e^{- V [a^2/(4b) +|\lambda| \sqrt{a/b} + \lambda^2/(4a) + \cdots]}
; \quad |\lambda| \ll 1,~a>0
\label{asy}
\eea
where $ z, ~ \phi $ are constants.

For the case of no spontaneous symmetry breaking ($a>0$) the free energy 
$F$ is well defined and satisfies (\ref{bound}) though it is not 
analytic in $ \lambda $.
When the symmetry is spontaneously broken the free energy suffers from
the ailments described in~\cite{a.g}. Note that the bound (\ref{bound}) is
satisfied in either case.

Since the free energy is ill defined when $ a < 0 $ (though (\ref{bound})
is still obeyed), the VW argument is incomplete in this case. The problem
is that the partition function oscillates, 
corresponding to a complex $W$; we now argue
that this will not occur in any physical system. Indeed,
a reasonable requirement for a system is that the Hamiltonian be an additively
renormalizable operator that is bounded from below. In this case
$\exp(-W)$ equals the $ \beta \to 0 $ limit of the partition function $
Z = \tr \; e^{ - \beta H } $. For
for any finite $ \beta $, the standard expression for $Z$
in terms of energy eigenvalues shows that it cannot be negative, 
provided we assume that the exponential decrease
in the contributions from large (renormalized)
energy states is sufficient to guarantee
convergence of the trace. Systems obeying these restrictions 
will have $ Z > 0 $, which then implies $ \exp(-W) \ge 0 $.
In such cases the possibility of negative  $ a $
(or any set of parameters leading to a complex $W$), where 
parity is spontaneously broken, will not be realized.

This same issue was discussed in~\cite{j}, where it is claimed that 
$W(\lambda )$, when defined appropriately, has the same dependence on
$ \lambda $ independent of the sign of $a$. Although we find that this is
not the case,
the claims of this reference would be validated by (\ref{asy})
provided the free energy is a concave function of
$\lambda $, as is indeed the case when $ a>0$. 

This, however, is not the case $ a<0$.
If we adopt the prescription of~\cite{j} of defining the free energy
through an average over the volume, the results are ambiguous. For example
the free energy per unit volume,
\beq
f = -\lim_{V \to\infty} \inv V \ln \ical
\eeq
would equal $ f = a^2/(4b)+\lambda^2/(4a) + \cdots $ provided we {\em define}
\beq
\lim_{V\to\infty} {\ln \cos( c V)\over V} =0 
\eeq
for any fixed real constant $c$. In this case the conclusions of~\cite{j}
follow: we cannot use the $\lambda$ dependence of the free energy to 
determine the presence or absence of symmetry breaking. However this is 
an {\it ad-hoc} choice: consider $ V_n = (n+1/2)\pi -e^{-u \pi n} $ for a real
constant $u$ and a positive integer $n$. Then letting $ V = V_n $ and taking 
the infinite volume limit by letting $n \to \infty $, we find
\beq
\lim_{n\to\infty} {\ln \cos V_n \over V_n } = i - u,
\eeq
having taken
the branch of the logarithm such that $ \ln \cos V_n = n \pi ( i -u )$.
With these choices the limit is certainly non-zero.
The point is that the large volume limit of $ (\ln \ical)/V $
is ill defined.\footnote{For a different
choice of the branch cuts the imaginary term would be absent.}
It is possible to define the limit so that the troublesome oscillatory
term does not contribute, but it is unclear whether this is the
physically correct definition of the free energy.

\medskip

We conclude that the concerns raised in~\cite{a.g,j} are valid in that
the partition function would not be positive definite in a system exhibiting
spontaneous parity violation. However, this  situation will not arise 
for systems with a  Hamiltonian bounded from below such that
the density of states
does not grow faster than the exponential of the energy.

\section{Lattice considerations}
\label{sect:lattice}

In a series of publications\cite{aoki} Aoki proposed a phase diagram
for lattice QCD that is markedly different from the standard case
(both quantitatively and qualitatively) for non-zero lattice
spacing (all such peculiar effects disappear in the continuum limit).
In this scenario the theory has two phases, one where the 3 pions
are massless, being the Goldstone bosons produced by the spontaneous 
breaking of the usual chiral symmetry. In the second phase
flavor symmetry is believed to be spontaneously broken down to
$U(1)$ generating two Goldstone bosons (for the case of 2 massless
quarks) so that the third pion is massive in this phase. The transitions
between phases are of the second order so that the third pion mass
vanishes smoothly as the phase boundary is approached. In this second
phase parity is also spontaneously broken leading to a non zero expectation
value of the order parameter $ \langle \overline{\psi} i \gamma_5 \psi \rangle $.

Though Aoki's original arguments were made within the lattice-regulated theory,
the results of Sect.~\ref{sect:proof} would be equally applicable to
this case\cite{ss} provided the various assumptions made are satisfied.
Should this be the case, the presence of a parity-violating phase
at finite lattice spacing casts doubt on the results of~\cite{vw},
even though the parity-asymmetric phase disappears in the continuum.

The lattice arguments in ref.~\cite{aoki} are supported by two large-N calculations, one for the
2-dimensional Gross-Neveu model\cite{gn} and one for 4-dimensional 
QCD\cite{4dQCD}\chg{; both
are based on finding minima for the effective potential that do not respect
parity. In
addition there is also some numerical evidence based on the Monte-Carlo
evaluation of several observables. We will revisit these calculations in the paragraphs below.}

\subsection{The lattice Gross-Neveu model}

This model contains $N$ spinor fields $ \psi_i $
where the doubling problem is fixed using Wilson fermions.
The dynamics is defined by the lattice action (we take a unit lattice spacing)
\bea
S &=& \inv{2} \sum_{i=1}^N 
\overline{\psi_i(n)} \left\{ 
\left[ \not\!\partial \psi_i \right](n) + \half r \left[ \square \psi_i
\right](n) + \left[ \sigma(n) - i \pi(n)\gamma_5 \right]\psi_i(n) \right\}
+ { N\over2g^2} \sum_n \left[ \sigma(n)^2 + \pi(n)^2 \right],
\label{GNaction}
\eea
where $g$ is $N$-independent and the derivatives denote the usual lattice
operations\cite{rothe}. 
In the large $N$ limit and for constant $\sigma $ and $ \pi $,
the fermionic determinant yields a function which has a minimum for a
non-zero value of $ \pi $. Even though quantum fluctuations are
known to destroy this non-trivial vacuum\cite{no.ssb}, this calculation
does suggest the possibility of the richer phase diagram described in~\cite{aoki}.

These results for the Gross-Neveu model do not contradict general argument
presented in~\cite{vw}. This is because this model
does not satisfy an important assumption made in deriving the 
VW result: since the fermions do not couple vectorially, the
fermionic determinant associated with (\ref{GNaction}) is not positive
definite. While this model can be used as guidance in obtaining
the phase structure of QCD on the lattice, it does not provide
a violation of the VW results.

\subsection{Lattice QCD in 4 dimensions at large $N$}

Another example considered by Aoki is the case of the large $N$ limit ($N$
again denotes the number of fermions) in lattice QCD in 4 dimensions using
Wilson's formalism. 
Denote by $ \sigma(n) = \psi(n)\otimes \overline{\psi(n)} $, then, following~\cite{aoki},
we consider the  effective potential for these operators in the case where they
are position-independent in the large $N$ and strong coupling limits. Replacing then
\beq
\sigma(n) \to \sigma e^{i \theta \gamma_5 }
\eeq
the effective action for this order parameter is (we take again
unit lattice spacing)\cite{4dQCD}
\beq
S_{\rm eff} = 
4 N  \Omega  \left\{ m \sigma \cos(\theta) - \ln(\sigma) - 
        2 \left[ \sqrt{1 - 4 \sigma^2 \sin(\theta)^2} - 1 - 
              \ln \left( {1 +  \sqrt{1 - 4 \sigma^2 \sin(\theta)^2} \over2} \right) 
\right], \right\}
\eeq
where $\Omega $ denotes the lattice volume, $m$ the bare fermion mass, and we have 
chosen the Wilson fermion parameter $r=1$. The effective potential is 
{\em minus} the above
expression. We will also replace
$ N \ln \sigma = (N/2) \ln \sigma^2 $ and avoid a spurious imaginary
contribution to  $S_{\rm eff}$.
This expression has a series of local extrema at the points
\beq
 (i):~ \cos\theta=\pm1,~\sigma=\pm1/m.\qquad 
(ii):~ \cos\theta=\pm m \sqrt{{3 \over 16-m^2}},~\sigma=\pm\sqrt{{3 \over 16-m^2}}.
\eeq
Points \chg{$(ii)$} are local minima of $S_{\rm eff}$
for $m^2<4$ but are saddle points for $m>2$; 
$(ii)$ are saddle points for $m^2<16$ and unphysical otherwise. 
It is remarkable that the curvature of the effective
potential at points $(ii)$ is positive in the $ \theta $ direction,
corresponding to a positive pion mass parameter.

None of these points, however, represent the global minimum
of the effective action. In fact we have the behavior,
\bea
S_{\rm eff} & \to & + \infty, \quad \sigma \to 0 , \cr
            & \to & - \infty, \quad \sigma \to \infty, ~ \theta = \pi .
\eea
Hence points $(i)$ represent meta-stable states where parity is broken,
the stable vacuum at $ \sigma =0 $ is parity even. \chg{Note that the stability
of the $ \sigma =0 $ vacuum is determined solely by the term $ \ln\sigma $ and
may be altered by finite $N$ corrections.}

This still does not explain why the numerical evaluation of $\langle \overline{\psi} 
\gamma_5 \psi \rangle $ yields a non-zero result. This is understood by
the observation~\cite{aoki} that the Wilson action does not lead to a positive
fermionic determinant, which follows form the fact that the Wilson term,
does not anticommute with $ \gamma_5 $. It is possible to generalize
somewhat the results of \cite{rfd}, and to show that the Wilson action
leads to a positive fermionic determinant, but {\em only} when the fermions
carry a real representation of the gauge group~\cite{ew}, which is not 
the case for QCD. Hence the presence of  spontaneous parity violation
in QCD with Wilson fermions does not invalidate VW for finite lattice
spacing, as the theory does not comply with the assumptions made in 
\cite{vw}. On the other hand, the lattice results do indicate that 
$\langle \overline{\psi} \gamma_5 \psi \rangle \to 0 $ in the continuum limit, 
as demanded by VW. In this sense the results in~\cite{aoki} {\em support} 
those of \cite{vw}.

\section{Fermionic order parameters}
\label{sect:fermion}

In a calculation related to Aoki's, Sharpe and Singleton\cite{ss}
considered the case of QCD with two flavors to which they added a
source for a parity-violating order-parameter. Then they studied the
vacuum expectation value of this order parameter as the source strength
was set to zero. The fermionic part of the Lagrangian in Euclidean space
is
\beq
\lcal_{\rm ferm} = \overline{\Psi} \left( \not\!\! D_E + m + h e^{ - i \theta
\gamma_5 \tau_3 } \right) \Psi,
\label{p.v.ops}
\eeq
where $ \Psi $ denotes the fermion iso-doublet (assumed degenerate in
mass), and $ h$ the source strength. The Euclidean covariant derivative operator $
\not\!\! D_E $ is anti-Hermitian (in our conventions the Euclidean gamma
matrices are Hermitian\cite{v.n}), and the Pauli matrix $ \tau_3 $ acts on the
isospin variables.

The possibility of spontaneous parity violation can be studied
considering the expectation values of operators of the form
\beq
\ocal\up a_\Gamma = \overline{ \Psi^a} \Gamma \Psi^a, \quad a = 1,2 ;
\label{def.o.gam}
\eeq
where $a$ labels the isospin components and $ \Gamma $ is a linear
combination of the unit matrix, $i \gamma_5,~\tau_3 $ and $ i \tau_3
\gamma_5$. To obtain the general form of the expectation value,
\beq
\left\langle \ocal\up a_\Gamma \right\rangle = \tr \left\{ \Gamma \inv{
\not\!\!D_E + m + h \exp[- i \theta \gamma_5 (\tau_3)_{a a} ] } \right\},
\label{the.trace}
\eeq
(no sum over $a$) we follow~\cite{ss,b} and consider the above system
in a finite volume; we then expand the above trace
using the eigenfunctions of $ \not\!\!D_E $ which we denote by $ \varphi_n $.
Explicitly,
\beq
\not\!\!D_E \varphi_n = i \lambda_n \varphi_n,
\eeq
where we used the fact that the Euclidean Dirac operator is anti-Hermitian. As
usual~\cite{ag} we will treat the zero modes separately.

When $ \lambda_n \not=0 $ the mode $ \varphi_n $ is paired with $ \gamma_5
\varphi_n$ for which
\beq
\not\!\!D_E \left( \gamma_5 \varphi_n \right) = i ( - \lambda_n ) \left(
\gamma_5 \varphi_n \right),
\eeq
hence in the $ \{ \varphi_n, \gamma_5 \varphi_n \} $ subspace we have
\beq
\not\!\!D_E + m + h e^{\mp i \theta \gamma_5 } \to \pmatrix{ i \lambda_n + m
+ h \cos\theta & \mp i h \sin\theta \cr \mp i h \sin\theta & - i
\lambda_n + m + h \cos\theta } ,
\eeq
and we can replace $ \gamma_5 \to \tau_1 $ (denoting the usual Pauli
matrix). Then the corresponding contributions to the trace (\ref{the.trace})
are
\bea
\tr \left\{ {\bf1} \inv{ \not\!\!D_E + m + h e^{ \mp i \theta \gamma_5 }}
\right\} \to {2 ( h \cos\theta + m) \over \lambda_n^2 + \Lambda^2 } ; \qquad
\tr \left\{( i \gamma_5) \inv{ \not\!\!D_E + m + h e^{ \mp i \theta \gamma_5 }}
\right\} \to \mp {2 h \sin\theta \over \lambda_n^2 + \Lambda^2 };
\eea
where
\beq
\Lambda^2 = h^2 + m^2 + 2 h m \cos\theta.
\eeq

For the zero modes $ \lambda_n =0 $ we can choose $ \varphi_n $ to be an
eigenstate of $ \gamma_5 $,
\beq
\gamma_5 \varphi_n = \chi_n \varphi_n , \quad \chi_n^2=1,
\eeq
In this one-dimensional subspace we can replace $ \gamma_5 \to \chi_n $
and
\beq
\not\!\!D_E + m + h e^{\mp i \theta \gamma_5 } \to h e^{\mp i \theta \chi_n
} + m .
\eeq
Then the corresponding contributions to (\ref{the.trace}) become
\bea
\tr \left\{ {\bf1} \inv{ \not\!\!D_E + m + h e^{ \mp i \theta \gamma_5 }}
\right\} \to { h e^{\pm i \theta \chi_n } + m \over \Lambda^2 }; \qquad
\tr \left\{ ( i \gamma_5) \inv{ \not\!\!D_E + m + h e^{ \mp i \theta \gamma_5 }}
\right\} \to i \chi_n { h e^{\pm i \theta \chi_n } + m \over \Lambda^2 } .
\eea

Adding all terms  we find
\bea
\tr \left\{ {\bf1} \inv{ \not\!\!D_E + m + h e^{ \mp i \theta \gamma_5 }}
\right\} &=& \sum_n { h \cos\theta + m \over \lambda_n^2 + \Lambda^2 } 
\pm i \nu { h \sin\theta \over \Lambda^2 } ,\cr
\tr \left\{(i \gamma_5) \inv{ \not\!\!D_E + m + h e^{ \mp i \theta \gamma_5 }}
\right\}  &=& \mp \sum_n { h \sin\theta + m \over \lambda_n^2 + \Lambda^2 } 
+ i \nu { h \cos\theta + m \over \Lambda^2 }, 
\eea
where the sum is over {\em all} modes (with positive negative and zero
eigenvalues), and $ \nu $ denotes the index of $ \not\!\!D_E$,
\beq
\nu = \sum_n \chi_n,
\eeq
where this sum is over the zero modes only.

Using these expressions we find, for the operators in (\ref{p.v.ops}),
\bea
\left\langle \ocal\up a_{\bf1} \right\rangle &=& { h \cos\theta + m
\over \Lambda  \Omega } \sum_n { \Lambda \over \lambda_n^2 + \Lambda^2 }  - i
(\tau_3)_{a a} { \nu\over\Omega}  { h \sin\theta \over \Lambda^2 } , \cr
\left\langle \ocal\up a_{i \gamma_5} \right\rangle &=& (\tau_3)_{a a} {
h \sin\theta \over \Lambda \Omega } \sum_n { \Lambda \over \lambda_n^2 + \Lambda^2 }
+ i { \nu\over\Omega}  { h \cos\theta + m \over \Lambda^2 },
\eea
(no sum over $a$) where $ \Omega $ denotes the space-time volume. In particular,
\beq
\left\langle \int d^4 x\; \overline{\Psi} i \gamma_5 \Psi \right\rangle = 2 i 
\nu { h \cos\theta + m \over \Lambda^2 }
\eeq
Taking the infinite volume limit and then letting $h$ approach zero, we
obtain
\bea
\left\langle \int d^4 x\; \overline{\Psi} i \gamma_5 \Psi \right\rangle_{h=0} 
&=& 2 i { \nu \over m } \cr
&=& {i \over  8 \pi^2 } \int d^4 x \tr \left[ \tilde F_{\mu\nu} F_{\mu\nu}
\right]
\label{p.v}
\eea
The authors of~\cite{ss} obtain zero instead of the above expression
since they did not include the contribution from the zero modes
explicitly. Note also that both isospin components contribute equal
amounts to the right-hand side of (\ref{p.v}). The fact that we obtain 
a non-vanishing expression in (\ref{p.v}) does {\em not} imply that parity is broken:
this will be true only if this non-zero value survives the integration
over the gauge fields. That integral cannot be evaluated, even formally;
yet the fact that the \rhs\ of (\ref{p.v}) is purely imaginary allows us
to use the argument leading to (\ref{bound}) which implies that, in
fact, parity is not broken.

It is worth noting that (\ref{p.v}) is consistent with the anomaly
equation in Euclidean space provided we assume that there are no
massless excitations, so that the volume integral of the divergence
of the axial current vanishes. We also note that the corresponding
expression for the parity-even bilinear gives
\beq
\left\langle \overline{\Psi} \Psi \right\rangle_{h=0} 
= - \int d\lambda \tilde\rho_A(\lambda ) { 2 m \over \lambda^2 + m^2 }
\label{ch.v}
\eeq
where $ \tilde \rho_A$ denotes the spectral density per unit
four-volume for the operator $ \not\!\!\! D_E$. This expression
reproduces the corresponding result of~\cite{ss}. 

If we add to the Lagrangian a term $ \theta \tr F \tilde F/(4\pi)^2 $
(where $F$ denote the field strength of the gauge field in the fermionic
covariant derivative and $ \tilde F$ the corresponding dual) 
then (\ref{p.v}) can be interpreted as stating
that the expectation value on the left-hand side can be compensated by
an appropriate shift in $ \theta $. For the case of low-energy
QCD this is precisely the same result obtained using a chiral
Lagrangian: a constant shift in the $ \eta'$ field can be compensated
by a shift in $ \theta $ \cite{gl}.

The expression (\ref{p.v}) validates the claims of~\cite{vw} in that the
expectation value is purely imaginary. We also note that (\ref{p.v}) is
ill defined in the massless case where it has a $ 1/h$ singularity. This
problem is connected to the fact that the operators under consideration,
$ \ocal\up a_{i \gamma_5 } $ violate {\em both} parity and chiral
symmetry. We can repeat the calculations for an operator that violates
only parity such as
\beq
\ocal\up a_D = \overline{\Psi^a} i \gamma_5 {\buildrel \leftrightarrow \over
{\not\!\!D}} \Psi^a,
\label{def.o.D}
\eeq 
In this case we find
\beq
\left\langle \ocal\up a_D \right\rangle = - i (\tau_3)_{a a} h
\sin\theta \int d\lambda \tilde\rho_A(\lambda ) { 2 \lambda \over \lambda^2 + m^2 },
\eeq
that vanishes as $ h \to 0 $, but also vanishes when $ \theta =0 $ when
the theory is parity symmetric. For this operator the expectation value
is zero at the level of the fermion integral, in contrast with
(\ref{p.v}).

It is worth pointing out that $\ocal\up a_D$ is {\em not} related to 
$ \ocal_{i\gamma_5}\up a$ in (\ref{def.o.gam}) through the use of the equations
of motion, in fact, if we replace $ \not\!\! D \to \not\!\! D +i m $
in (\ref{def.o.D}) $ \ocal\up a_D$ remains unchanged. This shows that 
this operator can be eliminated by performing a chiral rotation on the
fermion fields, and is
then equivalent to a purely bosonic operator proportional to the
index of the Dirac operator $ \not\!\! D_E $~\cite{fuj}.

The authors of~\cite{ss} also consider the behavior of the lattice
version of this theory close to the continuum. They find that the
leading contributions to the effective potential for the meson field
$ \Sigma = \sigma + i \taubf \cdot \pibf, ~ (\sigma^2 + \pibf^2 = 1)$,
equals
\beq
V_{\rm eff}( \Sigma ) = - c_1 \sigma + c_2 \sigma^2,
\eeq
with $ c_2 $ approximately constant (independent of $m$) while
$ c_1 $ is expected to be a linear function of $m$. The crucial issue
here is the sign of $ c_2 $: if $ c_2 < 0 $ the potential has a minimum
at $ \sigma^2 = 1 $ that corresponds to a parity-symmetric
vacuum where the pion field 
gets no expectation value. If, on the other
hand, $ c_2 > 0 $ parity will be broken~\footnote{In addition, the
phase structure suggested by Aoki is reproduced by this model when $ c_2
> 0 $.}. In view of the above discussion that supports the results of
Ref.~\cite{vw}, and suggests that the parity-violating vacua of the
large-N, strong coupling lattice theory are not absolutely stable, we
believe that, in fact, $ c_2 < 0 $ as well.

\section{Parity violation at finite temperature}
\label{sect:temp}

In~\cite{cohen} it is noted that systems in a heat bath have 
an available additional time-like vector $ n^\mu $ (the temperature vector), 
that can be used to construct P-violating operators that apparently 
do not acquire the crucial factor of $i$ upon a Wick rotation.
An example is the operator
$ \epsilon^{\alpha \beta \gamma\delta} n_\delta
\tr\left[ D_\alpha F_{\beta \nu} n^\nu F_{\gamma \mu} n^\mu \right] $.

In order to examine this claim we consider the case of time-dependent
field theory at finite temperature. In this case it is well known\cite{n.s}
that in the functional integral exponent, the time integration should 
be carried along a complex path starting at $ - \infty $ and ending at 
$ - \infty - i \beta $. The temperature vector is, in the rest
frame for the system, the tangent to this path, and so it is, in general,
complex. For the purposes of evaluating the expectation value of an
order parameter the path can be taken parallel to the imaginary axis and,
as a result, $n^\mu$ is purely imaginary. Since all
operators in the class mentioned above are odd in $ n^\mu$,
they {\em do} receive a factor of $i$ and will not
violate the claims of~\cite{vw}.

\section{Comments and Conclusions}
\label{sect:conclusions} 

The above calculations provide a re-examination of the original
Witten-Vafa result concerning the absence of spontaneous parity violation
in theories with only vector-like fermion couplings. We also
reviewed the various criticisms of that result and concluded that
none of them is strong enough to invalidate the conclusions of~\cite{vw}.

Theories with scalars can evade the VW result in two ways. Firstly
some pseudoscalar fields (if present) can acquire an expectation
value, but such order parameters do not receive a factor of
$i$ when Wick rotated. Secondly, in such theories
the fermionic determinant is not necessarily positive definite 
({\it e.g.} in theories containing solitons.\cite{jr}).

In confining and  parity-conserving theories containing fermions 
and gauge bosons, the low-energy excitations are often scalars 
and fermions (as is the case for
QCD). The comments above suggest that the low-energy effective
theories for such models can, in principle, violate parity
spontaneously. In fact the standard chiral non-linear sigma
model\cite{gl}, with the addition of an electromagnetic mass\cite{dp},
 allows parity-violating vacua for a certain range
of parameters. On the other hand, the VW argument implies  
that there is no parity violation in the underlying theory.
Of course, one could simply argue that the VW result for the underlying
theory merely requires that the low-energy parameters are such that
no parity violation occurs. Still, this situation
opens the possibility that there might be some non-perturbative
effects that allow such theories to evade the results of VW. An 
investigation of this possibility lies beyond the scope of 
the present publication, though we intend to discuss it fully in
the near future~\cite{ew}.

Finally, we note that, as briefly mentioned in~\cite{vw}, the same arguments
cannot be used to rule out spontaneous CP violation in theories
without scalars as there are CP-violating operators that remain real upon
Wick rotation (e.g. $ F_{\mu\nu}^a F_{\nu\rho}^b F_{\rho \mu}^c f_{abc} $
where $F$ denotes the non-Abelian field strength, and $f$
the group structure constants). So, even though it is known
that such theories cannot violate CP explicitly,\cite{cp} the possibility
of their exhibiting spontaneous CP violation remains.

\begin{acknowledgments} 
J.W. would like to thank T. Onogi for
his hospitality at the Yukawa Institute of Theoretical Physics
where part of this work was complted, and for
helpful comments concerning the Aoki phase. This work was supported
in part through funds provided by the U.S. Department of Energy under
grants DE-FG02-95ER40899 and DE-FG03-94ER40837.
\end{acknowledgments}

\section{Appendix}

In  this appendix we evaluate the integral
\beq
\ical(q) =  \int_{-\infty}^\infty dx\;
\exp\left[ - \inv4 x^4 - \alpha q^{2/3} x^2 + i q x \right]
\eeq
in the limit of large $ |\alpha| $.  Our strategy will be
to construct a differential equation for $ \ical $
whose solutions can be obtained
by a method similar to the WKB approach. The disadvantage of this
approach is that the differential equation has solutions that
do not correspond to $ \ical $. In order to extract the
relevant ones we will use the following property,
\beq
|\ical | \le  \int_{-\infty}^\infty 
\exp\left[ - \inv4 x^4 - \alpha q^{2/3} x^2 \right]
=e^{\alpha^2  q^{4/3}/2} \sqrt{|\alpha|  q^{2/3}}
K_{1/4}\left(\alpha^2  q^{4/3}/2 \right)
{\buildrel q \to \infty \over \longrightarrow} 
{ \sqrt{\pi/\alpha}\over 2^{1/4} q^{1/3}}
\label{condition}
\eeq
that follows from the Schwartz inequality.

Integrating by parts, we find (a prime denotes a $q$ derivative)
\beq
\ical'''+ u \ical''+ v \ical' + w \ical  =0 
\label{i.equation}
\eeq
where
\bea
u &=& -{8\over3} \alpha^2 q^{1/3} +
{8 \alpha \over 6 \alpha q - (27 - 32 \alpha^3)q^{7/3} } \cr
v &=& {4\over9} \alpha \left( 4 \alpha^3 -9 \right) q^{2/3} -{8\over9} \alpha^2 q^{-2/3}
- {(224/9) \alpha^3 q^{1/3}\over 6 \alpha q - (27 -32 \alpha^3)q^{7/3}} \cr
w &=& {4\over9}\alpha^2 q^{- 5/3}-{4\over3}\alpha q^{-1/3}+{16\over27}\alpha^4 q^{-1/3}
+\left({16\alpha^3\over27}-1\right)q - {32\over9}
{\alpha^2(\alpha q^{-2/3}+3 q^{2/3}) 
\over 6 \alpha q - (27 - 32 \alpha^3)q^{7/3}}
\eea

In the large $q$ limit the solutions for $ \ical$ can be obtained by using the Ansatz
\beq
\ical = \exp\left( - y q^{4/3} \right)
\eeq
that solves (\ref{i.equation}) for large $q$ provided
\beq
64 y^3 - 128 \alpha^2 y^2 + 16 \alpha(8\alpha^3-9) y+(16\alpha^3-27) =0 
\eeq

Of the 3 solutions to this equation one
is real for all $ \alpha $; we call it $ y_0 $. The two others are complex 
conjugate of each other for $ \alpha \lesim 0.944941 $ and real for 
$ \alpha > 0.94491 $; we denote them by $ y_\pm $ with the sign associated 
with the sign of the imaginary part. The asymptotic behavior of these solutions is
given in Table ~\ref{tab1}.

\begin{table}[h]
\begin{tabular}{|c|c|c|} \hline
solution & $ \alpha \to +\infty $ & $ \alpha \to -\infty $ \\ \hline
$y_0$ & $ \alpha^2 + \sqrt{2\alpha} + 1/(8\alpha) + \cdots $
    & $ - 1/(4\alpha) + \cdots $ \\ \hline
$y_+$ & $- 1/(4\alpha) + \cdots $
    & $ \alpha^2 + i \sqrt{2|\alpha|} + 1/(8\alpha)+ \cdots $ \\ \hline
$y_-$ & $ \alpha^2 +   \sqrt{2\alpha  } + 1/(8\alpha) + \cdots $
    & $ \alpha^2 - i \sqrt{2|\alpha|}  + 1/(8\alpha) + \cdots $ \\ \hline
\end{tabular}
\caption{Asymptotic behavior of the solutions to the differential equation
(\ref{i.equation})}
\label{tab1}
\end{table}

The required solutions should satisfy (\ref{condition}) which implies that the corresponding
$y$ should have a positive real part. In this way we find, for large $|\alpha| $ 
\bea
\ical &\simeq& Z e^{- q^{4/3}[\alpha^2 +1/(8\alpha) + \cdots]}
 \cos \left(\sqrt{2|\alpha|} q^{4/3} + \phi \right) ; \quad \alpha \to -\infty \cr
\ical &\simeq& Z e^{- q^{4/3}[\alpha^2 +\sqrt{2\alpha} + 1/(8\alpha) + \cdots]}
; \quad \alpha \to +\infty 
\eea
where $ Z, ~ \phi $ are constants. This is the result used in the text.

\end{document}